\documentclass[a4paper,10pt]{article}

\usepackage{amssymb}
\usepackage{amsmath}
\usepackage[usenames,dvipsnames]{color}
\usepackage{hyperref}
\usepackage[left=.8in,right=.8in,top=.8in,bottom=.8in]{geometry}              

\usepackage{xcolor}
\usepackage{mathpazo}
\usepackage[normalem]{ulem}
\usepackage{palatino}
\usepackage{mathrsfs}
\usepackage[mathscr]{euscript}

\usepackage{cite}
\providecommand{\BIBand}{\&}

\hypersetup{
    colorlinks=true,         
    linkcolor=MidnightBlue,          
    citecolor=BrickRed,        
    urlcolor=MidnightBlue             
}

\renewcommand{\tilde}{\widetilde}

\def\be{\begin{equation}}
\def\ee{\end{equation}}
\def\bea{\begin{eqnarray}}
\def\eea{\end{eqnarray}}

\def\fL{{\mathfrak{L}}}
\def\fI{{\mathfrak{I}}}
\def\d{{\mathrm{d}}}

\def\D{{\mathrm{D}}}

\def\scrF{{\mathscr{F}}}
\def\scrA{{\mathscr{A}}}

\renewcommand{\hat}{\widehat}
\renewcommand{\bar}{\overline}

\title{\bf Notes on a field-space connection}
\author{Henrique Gomes and Aldo Riello}
\author{{ Henrique \textsc{Gomes}\footnote{\href{mailto:gomes.ha@gmail.com}{gomes.ha@gmail.com}}}\,\, and { Aldo \textsc{Riello}}\footnote{\href{mailto:ariello@pitp.ca}{ariello@pitp.ca}}\vspace{.5em}
\\\normalsize Perimeter Institute for Theoretical Physics\\ \normalsize 31 N. Caroline St. Waterloo, ON, N2L 2Y5, Canada }
\date{\normalsize\today}

\begin{document}

\maketitle

\begin{abstract}
\noindent { We introduce a functional covariant differential as a tool for studying field space geometry in a manifestly covariant way. 
We then touch upon its role in gauge theories and general relativity over bounded regions, and in BRST symmetry. 
Due to the Gribov problem, we argue that our formalism---allowing for a non-vanishing functional curvature---is necessary for a global treatment of gauge-invariance in field space.  
We conclude by suggesting that the structures we introduce satisfactorily implement the notion of a (non-asymptotic) observer in gauge theories and general relativity.}

\end{abstract}

\vspace{1em}
\section*{Introduction}
Almost any introduction to gauge theory begins with a simple generalization, in which previously global symmetry transformations acquire  local spacetime dependence.
The extension is of the form  $g\rightarrow g(x)\in G$, where $x\in M$ is the point in the spacetime manifold $M$ and $G$ is the gauge group.
On account of the local dependence, terms such as  $\partial_\mu \overline \psi \partial^ \mu  \psi$ are no longer gauge invariant.
Thus arises the need for the introduction of a connection, which replaces the usual derivative by the covariant one, $\partial_\mu\rightarrow \D_\mu$, ensuring gauge \textit{co}variance of each derivative term. 

But in modern gauge field theory, one must take into account a much larger domain than spacetime, for the path integral is an integration over \emph{field space}, $\cal F$. 
Each element $\psi\in{\cal F}$ consists of one entire configuration of the field in question, $\psi=\{\psi(x)\}_{x\in M}$. Instead of performing an integration over $x$, as occurs in the classical action, we now must integrate over the infinite-dimensional $\cal F$.
The gauge group $G$ follows, being extended to the group of gauge transformations ${\cal G}=\times_{x\in M}G$.
In parallel to the previous generalization, $g\rightarrow g(x)$,  the extended transformations can \emph{a priori} also depend on the field configuration itself, i.e.  $g=\{g_\psi(x)\}_{x\in M}$.

These generalizations point to a natural question: should we also   extend our notion of covariant derivatives to the functional setting?
That is, we have a derivative on $\mathcal{F}$, called the functional derivative $\delta_\psi$, playing the role of $\partial_\mu$ in field space.
Shouldn\rq{}t we consider extending $\delta_\psi$ to a respective gauge-covariant functional derivative in the same vein as in $\partial_\mu\rightarrow \D_\mu$? 
The reason this question has not been deemed very consequential is simple. 
Clearly, it is not standard to consider gauge transformations that carry dependence on  the underlying field, i.e. gauge transformations of the form $g=g_\psi$. 
That is because their importance is obscure; in the classical regime we require no more than one field configuration, the equivalent of a single point $x$ in the finite-dimensional context, while the  non-perturbative path integral does not  by itself require any derivative on $\mathcal{F}$.\footnote{ The perturbative path integral on the other hand, may be seen as an integration over the tangent space at the point around which one is doing the perturbation. In this sense it does involve derivatives (or infinitesimals).}

Of course, such objects have not been completely ignored either, particularly in the canonical setting.  In the context of general relativity and related background independent gravitational theories,  non-purely-spatial diffeomorphisms were shown to generically fail to project onto the constraint surface, unless they have a specific phase-space dependence \cite{LeeWald}.
In \cite{Gauge_Riem}, one of us (HG) has defined functional connections in the context of a 3+1 gravitational theory.
In \cite{Conformal_geodesic}, this connection was used to define an action with a  horizontal projection (in the principal fiber bundle sense) for paths in configuration space. 

 However, the situation we investigate in these notes is  broader, since we shall consider gauge transformations which are field-configuration dependent already at the covariant (or presymplectic) level, i.e. before projection to the actual phase-space. Such connections need not be unique; they are required solely as a manner to make all quantities explicitly covariant in field space. However, essentially only one example of such a connection form has been studied in the literature---what we will here call the ``Vilkovisky connection''.
 
In \cite{vilkovisky1984gospel}, Vilkovisky introduced a functional connection within the covariant formalism,  in line with what we plan to discuss here. 
 DeWitt  \cite{DeWitt}  then built on Vilkovisky\rq{}s connection, showing that it removed the need for ghosts and argued for the usefulness of such objects in finding a renormalizable theory of quantum gravity.\footnote{HG thanks Steve Carlip for recently bringing this work to his attention.}  In \cite{Cotta-Ramusino-Reina}, a similar choice as that of Vilkovisky is mentioned (but not used).  More recently, Pawlowski applied the Vilkovisky--DeWitt framework to devise a diffeomorphism invariant RG flow in the background-field formalism, and thus probe the asymptotic safety scenario \cite{Pawlowski1,Pawlowski2}.
 Furthermore, various topological aspects of field theory, such as anomalies or some properties of the $\theta$-vacuum of QCD can be understood geometrically in terms of the introduction of a configuration-space connection and associated monodromies or curvature (e.g. \cite{WZ1971,zumino1984chiral}, and \cite{Wu}  for a more discursive account).

In this paper, however,  we will not be concerned explicitly with the path integral.  Instead, we will explore a surprising use of such connection forms even at a  classical setting, using  the covariant symplectic formalism on manifolds with corners (i.e.  on bounded regions). The main point is that gauge invariance in the presence of corners requires the introduction of such objects. 

In the context  of the (spacetime-)covariant symplectic formalism, the symplectic potential, can be written as a one-form in field space. For $S(\psi)$ the action functional related to the field configuration $\psi$ on the spacetime region $M$, we have: 
\be 
\delta S(\psi) = \int_M  \Big(\mathrm{EL}(\psi)\delta \psi+ \nabla_\mu \theta^\mu (\delta \psi,\psi)\Big)\epsilon
\ee
where $\nabla_\mu$ is a spacetime covariant derivative, $\mathrm{EL}(\psi)$ the Euler-Lagrange equations of motion, and $\epsilon$ the volume form on $M$.
Hence, for a codimension-one surface $\Sigma$, we get the presymplectic one-form on field space
\be
\label{symplectic_potential}
\Theta_{\Sigma}=\int _\Sigma \theta^\mu \epsilon_\mu ,
\ee
with $\epsilon_\mu$  the induced volume form of the surface. 
Note that $\Theta_\Sigma$ is called the (pre)symplectic potential whenever $\Sigma$ is a Cauchy surface. 
It is a one-form because $\delta$ is an exterior derivative in field space.
As we will show, this constitutes in our view a previously missed opportunity to employ a functional covariant exterior derivative.
 Most importantly, in spite of the  (pre)symplectic potential not being physically significant \emph{per se}, it is the precursor of fundamental physical quantities: the symplectic two-form, Noether charges and fluxes, as well as of (quantum) Aharonov-Bohm {and Berry} phases.

The main reason this potential use has been largely overlooked is that---as we will show---  the importance of functional covariant exterior derivatives  manifests itself mostly,  perhaps exclusively,\footnote{At least in the better understood case of Yang--Mills theories.} in the presence of corners, i.e. boundaries of $\Sigma$. 
Now, in the Cauchy case, it is most often  assumed that the spatial slices $\Sigma$ of the globally hyperbolic spacetime $M=\Sigma\times\mathbb R$ are compact. When this is not done, $\Sigma$\rq{}s boundaries are usually placed asymptotically far away, and  strong, gauge-breaking boundary conditions are imposed. 
For example, in gauge theories, all fields---and the gauge potentials among them---are required to vanish ``quickly enough'' at infinity. 
While in asymptotically flat gravity, the metric is required to be asymptotically Minkowski and diffeomorphisms are correspondingly reduced to their much more rigid counterparts---BMS transformations. 
    
But recently, the use of finite boundaries or less rigid asymptotic conditions \cite{barnich,campiglia16,Donnelly_2016, Hopfmuller,riello}  has been receiving more attention. 
In the context of finite boundaries, it was found \cite{Donnelly_2016} that to ensure covariance of the presymplectic potential new terms must be incorporated, with possibly profound implications for conserved quantities. 
In this paper, we will show that these extra terms emerge from the use of a functional connection one-form; it is a simple generalization of the finite-dimensional version to the infinite-dimensional setting. Interestingly, this formalism is also fully compatible with a geometric understanding of BRST, and, in fact, recovers it in the appropriate settings. 
    
We begin by briefly   recalling the derivation of the symplectic potential for gravity and Yang--Mills theories in the next section,  \ref{sec:Symplectic}. Then we review the most economical  description of gauge transformations through the use of principal fiber bundles, in section \ref{sec:PFB}.
This formalism introduces the notion of the gauge connection in a unified manner, translatable to the functional context, which we introduce next, in section \ref{sec:functional_PFB}.
In section \ref{sec:functional_symplectic}, we employ these tools to study a covariant symplectic potential for Yang--Mills theories and general relativity.
Then, in section \ref{sec_brst}, we discuss the geometric description of BRST, one of the standard formalisms to deal with gauge redundancy in field theories. In particular, we  show how our framework suggests a novel interpretation of this formalism in terms of a not-necessarily-flat connection one-form on field space. 
Lastly, in section \ref{sec:Gribov}, we comment on the relation with the Gribov problem, which makes the use of a non-flat connection over field-space an actual necessity.
In section \ref{sec:discussion}, we conclude the paper with a discussion on the physical interpretation of the formalism we introduced, which is  particularly pertinent for background-independent theories.   We suggest that the introduction of the field-space connection and of a proper notion of bounded regions can be used to encode the role of an observer in gauge theories and general relativity.

\section{Corners and field-dependent gauge transformations}\label{sec:Symplectic}

In this section we briefly review how the calculation of the symplectic potential is affected by the presence of corners in Yang--Mills theory and general relativity.

\subsection{Yang--Mills}\label{sec_YM}

From the Yang--Mills action for the curvature $\scrF[\scrA]$ of the $G$-connection $\scrA$ on {the spacetime manifold} $M$,\footnote{We omit the ad--invariant inner product on ${\frak g}=\text{Lie}(G)$, and---in most of the following---wedge products as well.}
\be
S_\text{YM} = {\frac12}\int_M \scrF\wedge\ast \scrF,
\ee
it is immediate to deduce the presymplectic potential of Yang--Mills theory: 
\be
\Theta_{\mbox{\tiny{$\Sigma$}}}(\scrA,\delta \scrA) = \int_\Sigma {E}\delta \scrA,
\ee
where $E =  \ast \scrF $ is the  electric field 2-form associated to the gauge potential $A$ (for definiteness in the handling of spacetime forms, we fix the spacetime dimension to $D=4$).
An infinitesimal gauge transformation $X(x)\in\mathfrak{g}$ acts as
\be
\left\{ 
\begin{array}{l}
E\mapsto E +\text{ad}_X E\\
\scrA\mapsto \scrA - \D_\scrA X := \scrA + \text{ad}_X \scrA - \d X
\end{array} 
\right. 
\ee
(We will from now on omit the subscript with the choice of surface, $\Sigma$, to avoid cluttering our notation.) 
This gives us the transformation of the presymplectic form, including a possibly field-dependent gauge parameter 
\be
\label{symp_form_YM}
\Theta \mapsto \Theta +  \int_\Sigma (\text{ad}_X E) \delta \scrA - E \D_\scrA\delta X + E (\text{ad}_X \delta \scrA) = \Theta - \int_\Sigma E \D_\scrA\delta X.
\ee
As we can see, in the absence of corners, the term that would include the variation of the gauge transformation vanishes after integration by parts due to the Gau{\ss} constraint,
\be
\D_\scrA E \approx 0.
\ee
In presence of a corner surface $C$, however, the resulting quantity is the electric flux across such a surface smeared against $\delta X$:
\be
Q^C_{\delta X}(\scrA)=  \int_C E \delta X.
\ee

\subsection{General relativity}\label{secGR}
  
In GR, we have a similar issue. 
In the following sections, we will pose the equations that show the same behavior in a more geometrical framework, but for now, we just follow the notation most common in the literature  (see e.g. \cite{Donnelly_2016}).

The Einstein-Hilbert action on the spacetime $M$ is 
\be
S_\text{EH} = \frac12 \int_M  R[g] \epsilon ,
\ee
where $R[g]$ is the Ricci scalar of the metric $g_{ab}$, and $\epsilon = \sqrt{|g|}\d x^0\wedge\dots\d x^{D-1}$ is the volume element on $M=\Sigma\times I$, $I\cong[0,1]$ being an interval.
The presymplectic potential for GR is then given by\footnote{Here $\delta g^{ab} := g^{aa'}g^{bb'} \delta(g_{a'b'}) = -\delta(g^{ab})$.}
\be
\label{symp_form_GR}
\Theta(g,\delta g) = \int_\Sigma \theta 
\qquad\text{and}\qquad
\theta = \frac{1}{2}\nabla_b\Big(\delta g^{ab}-g^{ab}(g^{cd}\delta g_{cd}) \Big)\epsilon_a,
\ee 
with (in form notation) $\epsilon_a = \iota_{\partial_a} \epsilon$ the induced volume element on the hypersurface $x^a=\text{const.}$.
An infinitesimal diffeomorphism is given by a vector field $X\in\Gamma(\mbox{T}M)\simeq\mbox{diff}(M)$,
\be
g_{ab} \mapsto g_{ab} + \pounds_{X} g_{ab} = g_{ab} + 2\nabla_{(a} X_{b)}.
\ee
Under such a transformation defined by a possibly metric-field dependent vector-field $X$, the presymplectic potential undergoes the change
\be
\label{symp_form_GR_gauge}
\theta \mapsto \theta + \pounds_X \theta + \nabla_b\Big(\nabla^{(a}\delta X^{b)}-g^{ab}\nabla_c \delta X^{c}\Big)\epsilon_a,
\ee
where $\delta X$ refers to the variation of $X$ with respect to its metric dependence.
By rearranging the term of \eqref{symp_form_GR_gauge} causing its non-covariance, we find that it is equal to
\be \label{symp_form_extra}
\nabla_b\Big(\nabla^{(a}\delta X^{b)}-g^{ab}\nabla_c \delta X^{c}\Big)\epsilon_a=
\left( G^a{}_b \delta X^b + \frac12 R \delta X^a  +\nabla_b \nabla^{[a}\delta X^{b]} \right)\epsilon_a %
 \approx \left(\frac12 R \delta X^a +\nabla_b \nabla^{[a}\delta X^{b]} \right)\epsilon_a,
\ee
with  $G_{ab} = R_{ab} - \frac{1}{2}R g_{ab}$ the Einstein tensor, and $\approx$ meaning equality on-shell of the constraint equations at (spacelike) $\Sigma$, $G^{ab}\epsilon_b|_{\Sigma} \approx 0$. Evaluated on a solution of vacuum GR, the non-covariant term above becomes the total derivative of a two-form and therefore does not contribute to $\Theta_\Sigma$ whenever $\partial \Sigma =0$. 
In presence of a corner surface $C=\partial \Sigma \neq \emptyset$, however, such a  two-form gives rise to the Komar integral associated to the vector field $\delta X$:
\be
Q^C_{\delta X}(g) = \int_C  (\nabla^{[a}\delta X^{b]}) \epsilon_{ab}.
\label{eq_komar}
\ee

This concludes our brief review of the required aspects of the presymplectic potential for gauge theories and gravity. We now move on to another necessary, and brief,  review---this time regarding the geometry of principal fiber bundles. 

  \section{Connections on principal fiber bundles}\label{sec:PFB}

The advantage of working directly with principal fiber bundles (PFB's) is that structures are simpler. 
For example, it is easier to work out particular features of associated vector bundles, or features of gauge-fixed structures directly from the general formalism of PFB\rq{}s than vice-versa.
The standard example of a PFB is the bundle of linear frames over a spacetime manifold $M$, with structure group $\mathrm{GL}(n)$.  
 
We will try as much as possible to adopt notation which can be naturally extended to the infinite-dimensional functional context. 

\paragraph*{PFB sections}

A principal fiber bundle $P$, is a smooth fiber bundle, for which a Lie group $G$ has an action from the right $G\times P \rightarrow P$,\footnote{This means that in the coordinate construction of $P$ the transition functions act from the left.} which we denote by $R_g p=: p\triangleleft g$, for $g\in G$, $p\in P$. 
The action is assumed to be free, so that $ p\triangleleft g=p $ iff $g=\mbox{id}_G$, the identity of the group.  
The quotient of {$P$} by the equivalence relation given by the group, that is $p\sim p\rq{}$ iff {$p'=p\triangleleft g$} for some $g\in G$, is  usually identified with the spacetime manifold, i.e. $M=P/G$.
We denote the projection map by $\mathsf{pr}:P\rightarrow M$. where $R_g$ denotes the right action of $g$ on $P$ (which should be distinguished from the action of $g$ on the group $G$ itself). The set $O_p=\{p\triangleleft g\, |\, g\in G\}$,  is called the orbit through $p$, or the fiber of $[p]=\mathsf{pr}(p)$. The action of $\mathsf{pr}$ projects $O_p$ to $[p]\in M$.

Let $U\subset M$ be an open subset of $M$, then a smooth embedding $\sigma:U\rightarrow P$ is a section (or a ``gauge-fixing'') if $\mathsf{pr}\circ \sigma = \mbox{id}_M$. 
The section then gives rise to a trivialization $G\times U\simeq \mathsf{pr}^{-1}(U)$, given by the diffeomorphism $F_\sigma:U\times G\rightarrow \mathsf{pr}^{-1}(U)$, $([p], g)\mapsto\sigma([p])\triangleleft g\,$.
 Generically, there are no such $U=M$ and the bundle is said to be non-trivial. {In the functional, i.e. field space, case, the base manifold is a modular space and  the obstruction to triviality is known as the Gribov ambiguity} \cite{gribov1977instability,singer1978some}.  We briefly review that in section \ref{sec:Gribov}.

\paragraph*{PFB connections}

Before going on to define a connection, we  require  the concept of a vertical vector in $P$. 
Let $\exp:{ \text{Lie}(G)=\mathfrak{g}}\rightarrow G$ be the group exponential map. 
Then by dragging the point with the group action we can define a vertical vector at $p$, related to $X\in \mathfrak{g}$, as 
\be 
\label{fundamental_vf} 
X_p^\#:= \frac{\d}{\d t}_{|t=0}\Big(p\triangleleft \exp{(tX)}\Big)\in \mbox{T}_pP
\ee
The vector field $X^\#\in \Gamma(\mbox{T}P)$ is  called the \emph{fundamental vector field} associated to $X$. 
The vertical space $V_p$ is defined to be the span of the fundamental vectors at $p$. 

From \eqref{fundamental_vf} one has that 
fundamental vector fields are ad--equivariant, in the following sense: 
\be
\label{ad_transf}
 ({R_g})_*X^\#_p=( \mbox{Ad}_{g^{-1}} X)^\#_{ p \triangleleft g}
\ee
where ${R_g}_*:\mbox{T}P\rightarrow \mbox{T}P$ denotes the pushforward tangential map associated to $R_g$ and $ \mbox{Ad}_g:\mathfrak{g}\rightarrow \mathfrak{g},~X\mapsto g X g^{-1}$ is the adjoint action of the group on the algebra. This notation emphasizes that the vector field at $p\in P$ is pushed-forward to $p\triangleleft g\in P$. Both sides of the equality, however, are evaluated at the same point, $p\triangleleft g \in P$.
Now, using the fact that the Lie derivative of a vector field $Y^\#$ along $X^\#$ is defined as the infinitesimal pushforward by the inverse of $\exp(tX)$ evaluated at $p$, by setting $g\to \exp(-tX)$ and $p \to p\triangleleft g^{-1}$ in \eqref{ad_transf} and deriving, we obtain
\be
\pounds_{X^\#} Y^\# = [X^\#,Y^\#]_{\mathrm{T}P}= \frac{\d}{\d t}_{|t=0}\Big({R_{\exp{(-tX)}}}_* Y^\# \Big)= \frac{\d}{\d t}_{|t=0} \Big({\mbox{Ad}_{ \exp{(tX)}}} Y \Big)^\#=(\mbox{ad}_X Y)^\# = ([X,Y]_{\frak g})^\#.
\label{eq17}
\ee
This shows that  the vertical bundle, $V\subset \mbox{T}P$, is an integrable tangential distribution. 

The definition of a connection amounts to the determination of an equivariant algebraic complement to $V$ in $\mbox{T}P$, i.e. an $H\subset \mbox{T}P$ such that
\begin{subequations}
\begin{align}
& H_p\oplus V_p=\mbox{T}_pP \\
&  {R_g}_*H_p= H_{  p \triangleleft g}\,,\, ~ \forall p
\end{align}
\end{subequations}
 One can equally well define $H$  to be the kernel of a $\mathfrak{g}$-valued one form $\omega$ on $P$, with the following properties:
\begin{subequations}
\begin{align}
 &  \omega(X^\#)=X   \label{omega-a}  \\
 & R_g{}^*\,\omega=\mbox{Ad}_{g^{-1}}\omega  \label{omega-b}
\end{align}
\label{omega}  
\end{subequations}
where $ {R_g}^*\omega\equiv \omega\circ {R_g}_*$. 
 In other words, equation \eqref{omega-b} intertwines the action of the group $G$ on $P$ (lhs of the equation) and its action on $\mathfrak{g}$ (rhs). Notice that these equations hold only for a global action of $G$ on $P$. 
For any $v\in\mbox{T}P$, its vertical projection is given by $\hat V(v)=\omega(v)^\#$. %
  
The horizontal derivative is the generalization of a covariant derivative (in the associated vector bundle) to the PFB context. For $\rho$ a representation of $G$ on the vector space $W$, and $\mu:P\rightarrow W$ an equivariant map, i.e. a map such that 
\be
(R_g^*\mu)(p) \equiv \mu(  p\triangleleft g)  =\rho(g^{-1})\mu(p),
\label{eq19}
\ee
the Lie derivative along a vertical vector field gives:
\be
\label{Lie_vertical}
 \pounds_{X^\#} \mu = -\rho(X) \mu,
\ee
where the corresponding representation of the Lie algebra on $P$ has been denoted by $\rho$,  as well. 
Also, applying twice Cartan's formula for the Lie derivative, i.e.
 \be
 \pounds_{X^\#} = \imath_{X^\#} \d + \d \imath_{X^\#},
 \ee 
and using $\d^2=0$, it is immediate to find that 
\be
\label{Commutation}
\pounds_{X^\#}\d =\d\pounds_{X^\#} .
\ee
Then, from this, and equation \eqref{Lie_vertical}, we can in turn deduce that for a point-dependent $X$ we have: 
\be
\label{non_equi}
 \pounds_{X^\#}\d \mu = -\rho(X)\d \mu - \rho(\d X) \mu
\ee
and thus $\d \mu$ is not equivariant.

 Regarding the previous formula, an abuse of notation has to be acknowledged (an abuse of notation that we will keep committing). By definition $X^\#$ is a ``global'' vector field encoding the global action of $G$ on $P$. Indeed, equations \eqref{eq19} and \eqref{Lie_vertical} have been introduced as reflecting the global action of $G$ on $P$. However, all formulas being local, they admit a straightforward generalization.  From \eqref{eq19} and \eqref{Lie_vertical}, we can extract local analogues,%
\footnote{In the following equation, an extra contraction with $\iota_{\hat V}$ should appear on the rhs if $\mu$ is generalized to also be a $p$-form $\lambda$, as in the following.}
\be
\pounds_{\hat V} \mu = - \rho(\omega)\mu
\ee
and
\be\label{Lie_d}
\pounds_{\hat V}\d \mu = \d \pounds_{\hat V} \mu,
\ee
where $\hat V$ is the projector on the vertical boundle $V\in{\rm T}P$.
Consider now a general vector field $v\in \mathrm TP $, we define their associated  field-dependent vector fields in $M$ by:  $X^\#_v  := \hat V (v) $ and $\omega(X^\#_v) := X_v$. Abusing notation slightly, we omit the label $v$ and work with point dependent vertical vector fields and point dependent Lie algebra elements, simply denoted as $X^\#$ and $X$, respectively.\footnote{Care must be taken, though, when using this notation; e.g. equation \ref{eq17} holds for fundamental vector-fields only, not for $X^\#_v$.} Consistently, we will write $\hat V = \omega^\#$. With this being understood, we go back to the non--equivariance of $\d\mu$.

To sort out that issue, one defines the horizontal derivative for general maps $\mu$ as:
\be
\label{hor_der}
\d_H = \iota_{\hat H} \d = \d - \iota_{\hat V} \d
\ee
 with $\hat H$ the horizontal projection in $\mbox{T}P$. For a zero form we have the equivalence $\d_H\mu=(\imath_{\hat H}\circ \d)\mu$. For an equivariant \emph{horizontal} form $\lambda$, i.e. for a $\lambda$ such that  $\pounds_{X^\#}\lambda =-\rho(X)\lambda$ and $\imath_{X^\#}\lambda=0$, we get
\be
\label{hor_hor} 
\d_H\lambda=  \d\lambda +\rho(\omega)\lambda .
\ee
 It is easy to check this formula: upon contraction with two horizontal  vectors the last term vanishes and equality is immediate. For contraction with a vertical vector,  the equality follows from the equivariance of $\lambda$ and equation \eqref{omega-a}. Indeed,
\be\label{Lie_equivariant}
\iota_{X^\#}(\d\lambda + \rho(\omega)\lambda) = \pounds_{X^\#} \lambda - \d \iota_{X^\#}\lambda + \rho(X)\lambda - \rho(\omega) \iota_{X^\#}\lambda 
 \ee
vanishes  for an horizontal equivariant form.

One then defines the Lie-algebra valued curvature 2-form as $F=\d_H\omega$. Since $\omega$ is not horizontal, one cannot use \eqref{hor_hor}, but instead gets%
\footnote{ The relation with the curvature used in the Yang--Mills action is $\scrF^\sigma= \sigma^* F$, where we emphasized the dependence of $\scrF$ on a choice of section $\sigma:M\rightarrow P$.}
\be
\label{curvature}
F=\d_H\omega=\d\omega +\frac{1}{2}[\omega,\omega]_\mathfrak{g}=-\omega([\hat H, \hat H]_{\mathrm{T}\it P})\,,
\ee
 where the first bracket denotes the Lie algebra commutator, while the second one the vector field commutator in $P$. This formula can be checked explicitly by inserting vertical and horizontal vectors. 
Let us e.g. contract $F$ with two vertical fundamental vector fields,\footnote{For general vector fields one can check that $F$ being a tensor the result of the contraction can depend only on the value of the tensor at the given point.} $X^\#$ and $Y^\#$: one first finds that $\iota_{Y^\#}\iota_{X^\#}\d \omega = X^\# Y - Y^\# X  - [X,Y] = - [X,Y]$ because $X$ and $Y$ are constant, and using $\iota_{Y^\#}\iota_{X^\#}[\omega,\omega] = 2[X,Y]$ we find the sought result. Notice that the case of two vertical vector fields is the only one which involves the term containing the commutator of two $\omega$'s. This will be relevant in the next section.

Equation \eqref{hor_hor} guarantees the equivariance of $\d_H\lambda$. From equations \eqref{hor_hor} and \eqref{Commutation}, and using the equivariance of $\lambda$ itself, we have 
\begin{align}
\pounds_{X^\#} \d_H\lambda & 
=\pounds_{X^\#} (\d\lambda+\rho(\omega)\lambda)=-\rho(\d X) \lambda -\rho(X)\d \lambda +\pounds_{X^\#}(\rho(\omega)\lambda).
\label{inhomo_Lie}
\end{align}
Now,
\begin{align}
\pounds_{X^\#}(\rho(\omega)\lambda)%
&=  \rho(\pounds_{X^\#}\omega)\lambda + \rho(\omega)(\pounds_{X^\#}\lambda)\nonumber\\
~&
=\rho(\d X)\lambda  - \rho( [X,\omega]_{\frak g})\lambda  - \rho(\omega)\rho(X)\lambda \nonumber\\
~& = \rho(\d X)\lambda - \rho(X)\rho(\omega)\lambda
\label{inhomo_Lie2}
\end{align}
 where in the first step we used the linearity of $\rho$ and that of the Lie derivative, in the second we used the equivariance of $\lambda$ and the following formula for the vertical Lie derivative of the connection, $\pounds_{X^\#}\omega = \d X - [X,\omega]_{\frak g}$. Such a formula can be deduced, e.g. from the horizontality of $F$, i.e. $\iota_{X^\#}F=0$, and equation \eqref{curvature}.
Thus, substituting in the rhs of \eqref{inhomo_Lie}, we obtain the sought equivariance property:
\be
\label{equi}
\pounds_{X^\#} \d_H\lambda= -\rho(X)\d_H\lambda.
\ee

\section{Functional connection on field space} \label{sec:functional_PFB}

Now, we will show how this structure is transferred to the infinite-dimensional, functional setting. 
We will focus on the case of general relativity, but an extension to Yang--Mills theories is immediate, and will be sketched at the end of the section.
We will ignore all the technical subtleties in transporting the constructions from section  \ref{sec:PFB} to the infinite-dimensional setting (for this point, at least in the Euclidean case,  see \cite{Gauge_Riem} and references therein).


\subsection{Standard functional setup} 

Let $\mathcal{F}$ be the space of metrics over $M$, it is the analogue of $P$ above.  
In the case of Euclidean metrics, this is the space of positive-definite sections of the symmetrized tensor bundle $\mathcal{F}=\Gamma_+(\mbox{T}M^*\otimes_S \mbox{T}M^*)$. 
Of course this is purely kinematical: no notion of equations of motion or gauge equivalence has been imposed on $\mathcal{F}$.

Importantly, we denote an element of $\mathcal{F}$ by $g$ (or $g_{ab}$ in the abstract index notation): this plays the role of $p\in P$. Furthermore, elements of $\mbox{Diff}(M)$ will be denoted by $\varphi$, and play the role of \emph{our former} $g\in G$.
Consequently, the Lie algebra of vector fields ${\frak X}^1(M)\cong\Gamma(\mbox{T}M)\ni X$, plays the role of $\mathfrak{g}$.
(Note that in this section, the symbol $g$ will stand for the metric configuration only, i.e. for a point of $\mathcal{F}$, and should not to be confused with the \emph{previous notation for a general group element}! On the other hand, elements of the Lie algebra have maintained their naming). 
The analogous to the previous base manifold, i.e. orbit space, is now the space of geometries, which we simply denote as $\mathcal{F}/\sim$, with the equivalence relation given by  $\mbox{Diff}(M)$.

A vector at $g\in\mathcal{F}$ is then tangent to a curve of metrics, and can be characterized through its action on functions on $\mathcal{F}$. I.e. for $v\in \mbox{T}_{g_o}\mathcal{F}$ such that\footnote{This makes sense because $\mathcal{F}$ is a linear functional space. Also, the tangent space to $\Gamma_+(\mbox{T}M^*\otimes_S \mbox{T}M^*)$ at any point is the linear Banach space $\Gamma(\mbox{T}M^*\otimes_S \mbox{T}M^*)$ with pointwise addition of the tensors. Furthermore, notice that the exponential map is also well-defined although not surjective onto any open set around $g\in \cal F$.} $\gamma=\frac{\d}{\d t}_{|t=0}(g_o+t\gamma)$, and for functionals $\mu(g)\in \mathbb{C}$: 
\be
\label{vec_action} 
\gamma(\mu(g))_{|g=g_o}=\frac{\d}{\d t}_{|t=0}\mu(g_o+t\gamma)=\int_M \d^D x \frac{\delta \mu}{\delta g_{ab}(x)}_{ |g=g_o}\gamma_{ab}(x)\,.
\ee
Similarly, we can define functional exterior derivatives, by first defining it on functionals, as $ (\delta \mu)(\gamma)=\gamma(\mu)~\forall \gamma$, and then using anti-symmetrization. From the usual definition, it follows Cartan's formula for the field-space Lie-derivative:
\be
\fL_\gamma=\delta\, \fI_\gamma +\fI_\gamma\, \delta,
\ee
where {$\fI$ is the analogue in field space of $\imath$, e.g. $\fI_\gamma\delta \mu = (\delta \mu)(\gamma) = \gamma(\mu)$}.
For a complete account on the infinite-dimensional aspects of differential geometry, see \cite{Lang} and, especially, \cite{Michorbook}.

As mentioned, what plays the role of the Lie group before is now a diffeomorphism $\varphi: M \to M$, $x\mapsto \varphi(x)$. It acts on the metric $g\in\mathcal{F}$ via pullback,%
\be
(A_{\varphi^*} g)_x := (\varphi^*g)_x.
\ee
Therefore, on functionals $\mu$ over $\mathcal{F}$, we have 
\be
(A_{\varphi^*}^*\mu)(g) = \mu( \varphi^*g),
\ee
and inifinitesimally, for a $X\in\Gamma({\rm T}M)$, %
\be
(X^\#\mu)(g) = \frac{\d}{\d t}_{|t=0} (A_{\Phi^*_{(tX)}}^*\mu)(g) =\frac{\d}{\d t}_{|t=0}\mu(\Phi_{(tX)}^*g)= \int \d^D x \; \frac{\delta \mu}{\delta g_{ab}(x)}\pounds_X g_{ab}(x),
\label{vertical_action}
\ee
where $\Phi_{(tX)}\in\mathrm{Diff}(M)$ is the flow of $X$: $\tfrac{\d }{\d t} \Phi_{(tX)}= X\Phi_{(tX)}$.
This is the analogue of \eqref{fundamental_vf}. For brevity we will sometimes write this in the form
\be
X^\#_g = \pounds_X g. \label{14analogue}
\ee

From these equations, in analogy with \eqref{ad_transf}, one deduces the following expression for the pushforward of the field space vector field $X^\#$ by the action of a diffeomorphism%
\footnote{See Appendix \ref{app}, for a detailed derivation. Contrary to equation \eqref{crucial}, and in analogy to equation \ref{ad_transf}, here we decided to emphasize that the field-space vector field has been pushed-forward from $g\in{\cal F}$ to $\varphi^* g\in{\cal F}$. Nevertheless, both sides of the equations are evaluated on the same field configuration $\varphi^* g\in{\cal F}$.}
\be
(A_{\varphi^*})_* X^\#_{g} = ( (\varphi^{-1})_*X\circ\varphi)^\#_{\varphi^*g}.
\ee 
Mimicking equation \eqref{eq17}, and taking the infinitesimal version of the above equality for $\varphi = \Phi_{(-tY)}$, we readily find $\fL_{Y^\#} X^\# = - (\pounds_Y X)^\#$, which can be re-expressed in the more suggestive form
\be
[Y^\#, X^\#]_{\mathrm T \cal F} = - ([Y, X]_{\mathrm T M})^\#.
\label{minus}
\ee 
This minus sign, absent from \eqref{eq17}, is a crucial technical detail. We will keep track of its consequences in the following.

For vector fields in $M$ which are metric dependent, i.e. $X:\mathcal{F}\rightarrow \Gamma(\mbox{T}M)$, we have $\delta X\in \Lambda^1(\mathcal{F},\Gamma(\mbox{T}M))$ -- a field-space one form valued in spacetime vector fields. 
Of course, using \eqref{14analogue} we can have the associated vertical vector field in $\Gamma(\mbox{T}\mathcal{F})$, i.e. $(\delta X)^\# \in \Lambda^1(\mathcal{F},\Gamma(\mbox{T}\mathcal{F}))$, where for $v\in \mbox{T}_g\mathcal{F}$,   $(\delta X (v))^\#\in \mbox{T}_g\mathcal{F}$ acts as in \eqref{vertical_action}.

For the analogue of \eqref{Lie_vertical}, generalize $\mu(g)$ from being a scalar, to being a spacetime differential form, $\mu(g)\in\Lambda^p(\mbox{T}M)$. That is $\mu(g)$ from now on acquires also a spacetime dependence, which we will leave however implicit.
Moreover, for simplicity, we restrict the representation to be simply the action by pullback\footnote{This is the analogue of the adjoint representation.} of diffeomorphisms on forms
{ Then, recalling that $(A_{\varphi^*}^*\mu)(g) = \mu(\varphi^* g)$, the equivariance condition for $\mu:\mathcal{F}\rightarrow \Lambda^p(\mbox{T}M)$ can be written as 
\be
\label{equivariant} 
 \mu(\varphi^*g)|_x = (A_{\varphi^*}^*\mu)(g)|_x  = \varphi^*(\mu(g))|_{\varphi(x)} ,
\ee
where on the lhs the diffeomorphism is acting in field space, while on the rhs on spacetime.\footnote{The rhs of this formula should {\it not} be confused with $A_{\varphi^*}^*\mu$! E.g., for $\mu(g)\in \Lambda^0(\rm TM)\cong C(M)$, we have $(\varphi^*\mu(g))_x = \mu(g)_{\varphi(x)}$.}}
Infinitesimally, we have the equivalent of \eqref{Lie_vertical},
\be
\label{inf_equivariant}   
\fL_{X^\#}\mu=\pounds_{X}\mu,
\ee
where $X$ is the vector flow of the infinitesimal diffeomorphism $\varphi$.
Two such transformations compose in reversed order,
\be
(A_{\psi^*}^* A_{\varphi^*}^*\mu)(g) = \varphi^*\psi^*\mu(g),
\ee
a fact consistent with the sign change in equation \eqref{minus}.\footnote{Explicitely: $(A_{\psi^*}^* A_{\varphi^*}^*\mu)(g) = (A_{\varphi^*}^*\mu)(\psi^* g) = \mu(\varphi^*\psi^* g) = \mu((\psi\circ\varphi)^*g) = (\psi\circ\varphi)^*\mu(g) = \varphi^*\psi^*\mu(g)$.
Yet another way of seeing the consistency of these formulas is the following. Suppose $X$ is field-space independent, then $\fL_{Y^\#}\fL_{X^\#} \mu = \fL_{Y^\#}\pounds_{X} \mu =  \pounds_{X} \fL_{Y^\#}\mu = \pounds_{X} \pounds_{Y}\mu $, since the spacetime and field-space differentials commute (and so do spacetime and field-space vector-field contractions).}

Such  $\mu:\mathcal{F}\rightarrow \Lambda^p(\mbox{T}M)$ are called $(0,p)$-forms, as they are  0-forms on $\mathcal{F}$ and $p$-forms on $M$.

{Note how the above equivariance condition is satisfied if and only if $\mu$ is a ``background independent'' quantity, that is a quantity that depends on spacetime only via dynamical fields belonging to ${\cal F}$. Indeed, if this were not the case, the right-hand sides of the two equations above would contain derivatives of fields not varied on their left-hand sides.}

Finally, we come to the analogue of the fact that $\delta \mu$ is not equivariant for a gauge-transformation that is point dependent, i.e.  $\fL_{X^\#}\delta \mu\neq \pounds_{X}\delta \mu$.  
Instead, for $\mu$ an equivariant $(0,p)$-form and $\beta:\mathcal{F}\to \mbox{Diff}(M)$ a diffeomorphism with field dependence,
using the equivariance condition \eqref{equivariant}, we find\footnote{Notice that we are leaving the spacetime dependence, i.e. the dependence on $x\in M$ of $\mu(g)_x$, implicit on both the lhs and the rhs of this and the following equations.}
\be
\delta (\mu(\beta^*g)) = \Big(\delta \mu +  \fL_{(\delta\beta\circ\beta^{-1})^\#}\mu \Big)(\beta^*g).
\label{eq_app2}
\ee
The calculation is performed explicitly in Appendix \ref{app2}.

For infinitesimal diffeomorphisms, this becomes
\be
\delta( \fL_{X^\#}\mu)= \pounds_{X}\delta \mu+\fL_{\delta X^\#}\mu\,.
\ee
Now, using Cartan's formula in field space to manipulate the last term, we obtain the exact analogue of \eqref{non_equi}:
\be 
\delta( \fL_{X^\#}\mu)=\pounds_{X}\delta \mu+\fI_{\delta X^\#}\delta \mu .
\ee
Similarly, using again Cartan's formula, we can rewrite the lhs of this equation in a form whose finite dimensional analogue is equation \eqref{Commutation},  that is
\be
\delta( \fL_{X^\#}\mu)=\delta(\fI_{X^\#}\delta \mu)=  \fL_{X^\#}\delta \mu.
\label{eq49}
\ee
Combining the last two formulas, we derive the infinitesimal statement of the non-covariance of the field-space Lie derivative
\be  
\fL_{X^\#}=\pounds_{X}+ \fI_{\delta X^\#}.
\label{hopf18}
\ee   
In this expression, we left understood that by linearity of the wedge product and the properties of the Lie-derivative, the above result can be extended to arbitrary $(n,q)$-forms. 

\subsection{Functional covariant derivative}  

In rough terms, the job of a covariant derivative is to exactly cancel out the non-equivariant term discussed in the last paragraph.    
We replace 
\be
\label{hor_field}
\delta\leadsto \delta_H= 
 \delta - \fI_{\hat V}\delta\;,
\ee 
where  $\varpi$ is a Lie-algebra-valued one-form, i.e. $\varpi\in\Lambda^1(\mathcal{F}, \Gamma(\mbox{T}M))$, which obeys {the analogue of }\eqref{omega}, that is
\begin{subequations}
\label{connection1and2}
\begin{align}
&\varpi(X^\#) =X\label{connection1}\\
&A_{\varphi^*}^*\,\varpi=(\varphi_*\varpi\circ \varphi^{-1}) \label{connection2}  
\end{align} 
\end{subequations}
and $\hat V$ is the vertical projection in $\mathcal F$. As before, the horizontal projection $\hat H$, is defined via the kernel of $\varpi$. At the price of a slight abuse of notation, we will write $\hat V = \varpi^\#$ (see discussion after equation \eqref{Lie_d})
Hence, for a field-dependent connection we have the analogue of \eqref{curvature} for the curvature:
{
\be
\mathfrak{F} = \delta_H\varpi = \delta \varpi - \frac12 [\varpi,\varpi],
\label{fieldspcurvature}
\ee 
where $[\cdot,\cdot]$ is just the usual commutator of vector fields on $M$ (Lie bracket), $ [\cdot,\, \cdot]_{\mathrm{T}M}$, and the unusual sign in front of the commutator is a direct consequence of equation \eqref{minus}. This can be shown e.g. by reproducing in the functional case the arguments put forward after equation \eqref{curvature} and recalling the difference between equations \eqref{eq17} and \eqref{minus}.

As in the PFB case, the vanishing of the curvature is equivalent to the nilopotency of $\delta_H$. We will explicitly verify this for an example in the next subsection.

If the curvature vanishes, we obtain the Maurer--Cartan equation
\be\label{BRST_omega1} \delta\varpi\;\hat=\frac{1}{2}[\varpi,\varpi],
\ee  
where the hatted equality means equality under the condition ${\frak F} = 0$.
However, we stress that even if the curvature does \textit{not} vanish, the following still holds
\be
\label{BRST_omega2}
\fI_{\varpi^\#}\delta \varpi=  \frac{1}{2}[\varpi,\varpi],
\ee     
being a direct consequence of the horizontality of $\frak F$. 
This relation will be required in sections \ref{sec5.1} and \ref{sec_brst}. 

Lastly, another statement equivalent to the horizontality of $\frak F$ that we will use in the following is
\be\label{Field_Lie_omega}
\fL_{X^\#}\varpi=(\delta\fI_{X^\#}+\fI_{X^\#}\delta)\varpi=\delta X + [X,\varpi].
\ee
Again, we stress the presence of a plus sign in the last term.

After the horizontal derivative of the field-space connection $\varpi$, the next natural object to study are $(n,p)$-forms.
For a $(0,p)$-form, we obtain
\begin{subequations}
\be
\label{BRST_A} 
\delta_H \mu=\delta \mu- \fI_{\varpi^\#}\delta \mu,
\ee 
where 
\be
\fI_{\varpi^\#}\delta \mu=   \int_M \d^D y\;\frac{\delta \mu}{\delta g_{ab}(y)}\pounds _\varpi g_{ab}(y).
\ee
\end{subequations}
Hence, combining these with equations \eqref{hopf18} and \eqref{connection1}, we obtain what we were aiming for: through the use of the horizontal functional derivative, we \emph{automatically implement general diffeomorphism equivariance in field space}:
\be
\label{hor_equivariant}   
\fL_{X^\#}\delta_H \mu=\pounds_{X}\delta_H \mu.
\ee   

The analogous of  \eqref{hor_hor} for $(n,p)$--forms is found by implementing the covariant derivative of \eqref{hor_field} on (field-space) {\it equivariant} and (field-space) \textit{horizontal} $(q,p)$-form $\lambda$, i.e. forms such that $A_{\varphi^*}^* \lambda = \varphi^*\lambda$ and $\fI_{X^\#}\lambda = 0$ respectively. The result then is 
\be
\label{hor_hor_field}
\delta_H \lambda=\delta \lambda - \pounds_\varpi\lambda.
\ee
As a simple exercise, we now explicitly check the equivariance of this expression:
\begin{align}
\fL_{X^\#}\delta_H\lambda 
&\hspace{-.5mm}
\stackrel{\eqref{eq49}}{=}  \delta \fL_{X^\#}\lambda - \pounds_{\fL_{X^\#}\varpi}\lambda - \pounds_\varpi\fL_{X^\#}\lambda\notag\\
&\hspace{-.5mm}\stackrel{\eqref{Field_Lie_omega}}{=} \delta \pounds_{X}\lambda - \pounds_{\delta X + [X,\varpi]}\lambda - \pounds_\varpi\pounds_{X}\lambda
%
= \pounds_X \delta_H\lambda
\end{align}
This is one of our fundamental equations. Its consistency with the action by pullbacks is ensured by the sign in equation \eqref{minus}, which crucially percolated through the definition of $\frak F$ and equation \eqref{Field_Lie_omega} into the above computation.

Note also that we do not need to require the vanishing of the curvature form for this to hold. The same is true in  the Yang-Mills case. 

Let us conclude this section with two remarks.
As in the finite-dimensional case, the existence of the above construction does not imply that we should just substitute $\delta$ wherever it appears by $\delta_H$: there are situations in which we are interested in the Lie derivative, and not the covariant one, or in the exterior derivative, not in the exterior covariant one. But it does provide an explicit and elegant implementation of gauge invariance in field space. 

Finally, recall the discussion after equation \eqref{inf_equivariant}, where we observed that the equivariance condition for a $(0,p)$-form encoded its ``background independence''. Here we stress the same point: the use of the field-space horizontal variation $\delta_H$ is the correct way of maintaining manifest spacetime background independence (and/or gauge covariance) in field space. In particular, a horizontal presymplectic form is one which identically vanishes when contracted with the Hamiltonian vector field associated to an arbitrary diffeomorphism transformation, even a field-dependent one. We turn now exactly to these matters, focusing our formalism towards  presymplectic potentials in general relativity and Yang-Mills theories.

\section{Gauge-invariant symplectic geometry}\label{sec:functional_symplectic}


\subsection{Yang--Mills}

Although we have so far focused on a formalism mostly suited to general relativity, we begin this section applying the formalism to Yang--Mills theory, because it is simpler and allows us to highlight some features of the formalism while avoiding certain delicate subtleties.

In the case of Yang--Mills theory, the functional 1-form $\varpi$ is an element of $ \Lambda^1({\cal F}, \mbox{Lie}(\mathcal G))$, where $\mathcal G = \times_{x\in M} G$ is the group of gauge transformations at a point of $\mathcal F$.
In particular, $\varpi$ is such that for any $X \in\mbox{Lie}({\cal G})$ and ${g}\in {\cal G}$ , one has
\begin{subequations}
\begin{align}
&\varpi(X^\#) =  X\\
&R^*_{{g}} \varpi = \mbox{Ad}_{g^{-1}}\varpi
\end{align}
\end{subequations}
And for a field-dependent gauge transformation $\beta : \mathcal F \to \mathcal G$, 
 \be
R_{\beta}^* \varpi = \mbox{Ad}_{\beta} \varpi - \delta \beta \beta^{-1} 
\ee
where, in the expression $\beta^{-1}$ the inverse is that in the group $\mathcal G$.

Using the fields defined in section \ref{sec_YM}, one readily finds the following expressions 
\begin{subequations}
\begin{align}
 \delta_H E = \delta E + [\varpi, E] \\
 \delta_H \scrA = \delta \scrA - \D_\scrA \varpi = \delta \scrA +[\varpi, \scrA] - \d \varpi \label{deltaHA}
\end{align}
\end{subequations}
are gauge covariant:
\begin{subequations}
\begin{align}
&\delta_H (\mbox{Ad}_\beta E) = \delta (\mbox{Ad}_\beta E ) + \mbox{Ad}_\beta[\varpi, E] - [\delta \beta \beta^{-1}, \mbox{Ad}_\beta E] = \mbox{Ad}_\beta \delta_H E\\
&\delta_H (\mbox{Ad}_\beta \scrA - \d \beta \beta^{-1}) = \delta (\mbox{Ad}_\beta \scrA - \d \beta \beta^{-1} ) + [\mbox{Ad}_\beta \varpi - \delta \beta \beta^{-1}, \mbox{Ad}_\beta \scrA - \d \beta \beta^{-1}] - \d (\mbox{Ad}_\beta \varpi - \delta \beta \beta^{-1})\notag\\
&\phantom{\delta_H (\mbox{Ad}_\beta \scrA - \d \beta \beta^{-1}) }
 = \mbox{Ad}_\beta (\delta \scrA + [\varpi, \scrA]) 
 - [\mbox{Ad}_\beta \varpi , \d \beta\beta^{-1}] 
 - \d (\mbox{Ad}_\beta \varpi)
  - \delta( \d \beta \beta^{-1})+ \notag\\%
  &\phantom{\delta_H (\mbox{Ad}_\beta \scrA - \d \beta \beta^{-1}) }\quad
 + [ \delta \beta \beta^{-1}, \d \beta\beta^{-1}] 
  + \d ( \delta \beta \beta^{-1})\notag\\
&\phantom{\delta_H (\mbox{Ad}_\beta \omega + \d \beta \beta^{-1}) }
= \mbox{Ad}_\beta (\delta_H \scrA) 
\end{align}
\end{subequations}
Moreover, it is straightforward to check that  the application $\delta^2_H$ to $E$ and $\scrA$ gives terms linear in\footnote{For Yang--Mills theories, the analogue of equation \eqref{minus} has the same sign as in the PFB section, essentially because the two derivations are tailored on the same algebraic structures. The sign propagates to the relevant formulas.\label{fntminus}} $\frak F$:
\begin{subequations}
\begin{align}
\delta^2_H E & = \delta (\delta E + [\varpi,E]) + [\varpi, \delta E + [\varpi,E]] =  [\delta\varpi, E] - [\varpi, \delta E]  + [\varpi, \delta E] + [\varpi,[\varpi,E]] = [\mathfrak{F},E]\\
\delta^2_H\scrA&  = \delta(\delta \scrA + [\varpi,\scrA] - \d\varpi) + [\varpi, \delta\scrA + [\varpi,\scrA] - \d\varpi] = [\delta\varpi,\scrA]  + [\varpi,[\varpi,\scrA]] - \d\delta \varpi - [\varpi,\d\varpi] = -\D_\scrA {\frak F}
\end{align}
\end{subequations}
Notice that in the second step, when $\delta$ ``crosses'' $\varpi$ it gets a minus sign, since we are permuting $(1,0)$--forms. On the other hand, the permutations of $(0,p)$ and $(n,0)$--forms comes with no extra sign. Accordingly, we also made use of the (graded) Jacobi identity:\footnote{The Jacobi identity holds because the $(n,p)$--forms, once evaluated at any point and contracted with any bivector in field space, take values in a Lie algebra. The signs are dictated by the $(n,p)$--form nature of the entries in the brackets.}
\be
0= [\varpi,[\varpi, E]] - [\varpi,[E,\varpi]] + [E,[\varpi,\varpi]] = 2[\varpi,[\varpi, E]] - [[\varpi,\varpi],E],
\ee
and similarly for $[\varpi,[\varpi,\scrA]]$.\footnote{In this appendix \ref{app_direct}, we present an explicit computation with $(n,p)$-forms.}

Now, recall the Yang--Mills presymplectic $(1,0)$--form
\be
\Theta(\scrA,\delta \scrA) = \int_\Sigma E\delta \scrA,
\ee
where in our notation the contraction by an ad-invariant form on $\frak g$ is left understood.
From the above formulas it is evident that the presymplectic potential calculated via the covariant exterior derivative is gauge invariant
\be
\Theta_H(\scrA, \delta\scrA,\varpi) = \int_\Sigma E\delta_H \scrA ,
\ee
and so is the symplectic $(2,0)$--form  
\be
\Omega_H = \delta_H\Theta_H =   \int_\Sigma  \delta_H E \delta_H\scrA  + E \delta^2_H \scrA,
\ee
Explicitly,
\begin{subequations}
\label{ymdeltaomega}
\begin{align}
\Theta_H &= \int_\Sigma E \delta \scrA + \varpi ( \D_\scrA E)  - \d E\varpi
\approx  \Theta -  \int_C E \varpi   
\\
\Omega_H &= \Omega +  \int_\Sigma\delta(\varpi \D_\scrA E) - \d \delta(E\varpi)
= \delta\Theta_H
 \approx \Omega -  \int_C \delta(E \varpi) ,
\end{align}
\end{subequations}
where $\approx $ stands for an equality on-shell of the Gau{\ss} constraint. 
The last equality, proved by a direct calculation in appendix \ref{app_direct}, could have been deduced from the following facts: $\Theta_H$ is gauge invariant, and $\delta_H$  coincides with $\delta$ on gauge invariant quantities, since $\delta_V\Theta_H=0$.
Notice that the extra terms which make the symplectic form and potential covariant are purely corner terms on-shell of the Gau\ss~constraint.

This symplectic potential and symplectic form in their most general form do not reproduce---to the best of our knowledge---any proposal previously advanced in the literature. The general form involves a non-vanishing  functional curvature, $\frak F \neq 0$. Were we to ignore the Gribov problem  (see section \ref{sec:Gribov}) and require $\frak F$ to vanish, then we could define a flat $\varpi$ connection as $\varpi:=\phi^{-1} \delta\phi$, or $\varpi=\delta_\phi$ in the notation of \cite{Donnelly_2016}, where $\phi:\Sigma\to G$ (on-shell of the Gau\ss~constraint, a choice $\phi:C \to G$ suffices). In this situation, we recover the same presymplectic form as the one proposed in \cite{Donnelly_2016}, but with a $\varpi$ defined also for non-vertical vectors. Notice that $\phi$ in this case is a {\it new} field whose addition  to the field space ${\cal F}$ is required by this procedure. This explains the name ``extended field space construction'' adopted in \cite{Donnelly_2016}.

Finally, notice that we have implicitly assumed that $\Sigma$ is given independently of the fields, $\delta\Sigma\equiv0$. 
This is equivalent to studying Yang--Mills theory on a fixed background, on the top of which a codimension-1 region $\Sigma$ is identified independently of the physical arrangement of the fields. In the next section, we  will discuss thoroughly how this assumption has to be rejected in the context of general relativity.

\subsection{General relativity\label{sec5.1}}

Replacing in the presymplectic $(1,0)$-form $\Theta$ of general relativity \eqref{symp_form_GR}, $\delta g_{ab}\leadsto\delta_H g_{ab}=\delta g_{ab}-2\nabla{}_{(a}\varpi_{b)}$,\footnote{This formula is easy to check from property \eqref{connection1}, $\mathcal{I}_{X^\#} \delta_H g_{ab}=0$.} we obtain 
\be
\theta_H(g,\delta g, \varpi)= { \theta(g,\delta_Hg) }= \frac{1}{2} \left[\nabla_b\Big(\delta g^{ab}-g^{ab}(g^{cd}\delta g_{cd})\Big) -\nabla_b\Big(\nabla^{(a}\varpi^{b)}-g^{ab}\nabla_c\varpi^c\Big)\right]\epsilon_a \,.
\ee
From the transformation properties of the connection form, it is clear that the extra term arising in \eqref{symp_form_GR_gauge},  i.e.
\be
\nabla_b\left(\nabla^{(a}\delta X^{b)}(g)-g^{ab}\nabla_c \delta X^{c}(g)\right)\epsilon_a,
\ee
cancels with the terms provided by the transformation of the field-space connection $\varpi$. Thus,
\be
 A_{\varphi^*}^*\theta_H 
 = \varphi^*\theta_H.%
\ee
This equation states the background independence and diffeomorphism covariance of  $\theta_H$  as a $(1,D-1)$-form.

To define an actual presymplectic form, rather than just a density, we need, however, to integrate $\theta_H$ over a codimension-1 hypersurface $\Sigma$,
\be
\Theta_H = \int_\Sigma \theta_H.
\ee
For this integration not to spoil background independence and diffeomorphism covariance, {\it $\Sigma$ must necessarily  depend on the field configuration}, so that
\be
A_{\varphi^*}^* \Sigma = \varphi(\Sigma).
\label{Sigma_covariance}
\ee
If this equation is satisfied, then $\Theta_H$ defines a fully diffeomorphism \textit{invariant} presymplectic $(1,0)$-form.  Again, it implies that $\delta_V\Theta_H=0$.

In other words, \emph{the integration region itself must be defined in terms of physical fields}, thus ensuring that the action of  diffeomorphisms  in field space do not alter the physical arrangement of fields and boundaries in the region of interest. This is far from a shortcoming; it is rather a manifestation of the relational nature of general relativity.

Such subtleties can be safely overlooked if $\Sigma$ is a closed hypersurface and the considered diffeomorphism is tangential to it (cf. equations \eqref{symp_form_GR_gauge}, \eqref{symp_form_extra}, and \eqref{eq_komar}). However, when considering a bounded hypersurface $\Sigma$, $\partial\Sigma\neq\emptyset$, or diffeomorphisms that are not automorphisms of $\Sigma$ itself, than all these considerations become crucial.\footnote{For these reasons, tangential diffeomorphisms of a closed Cauchy hypersurface behave essentially as actual (infinite dimensional) gauge symmetries, while this is not the case for transverse diffeomorpshisms \cite{LeeWald}.}

Having defined a fully invariant presymplectic potential, we can now move on and define a presymplectic form. Following our prescription, we define the covariant presymplectic form as
\be
\Omega_H:=\delta_H\Theta_H.
\ee
Nonetheless, thanks to the full invariance of $\Theta_H$, this is just equivalent to
\be
\Omega_H = \delta\Theta_H.
\label{normaldelta}
\ee
This fact is crucial for the following reason. 
To properly define the symplectic formalism of the theory, a nilpotent exterior derivative (one whose square vanishes)  and a closed (pre)symplectic form are indispensable features. Now, to keep manifest diffeomorphism covariance we have been forced to replace $\delta\leadsto\delta_H$. On the other hand, while the nilpotency of $\delta$ is guaranteed by definition, the nilpotency of $\delta_H$ is equivalent to the vanishing of the curvature of $\varpi$, i.e. $\delta_H^2 = 0 $ iff $\frak F = 0$. However, because of the Gribov problem (see section \ref{sec:Gribov}), $\frak F = 0$ can \textit{not} be imposed globally on field-space. Equation \eqref{normaldelta} allows us to circumvent this possible issue, since it shows that---as a consequence of its full diffeomorphism invariance---$\Omega_H$ is always $\delta$-exact. Hence, despite the necessity of introducing the horizontal $\delta_H$ to construct fully invariant quantities, once these are available, the diffeomorphism--invariant symplectic geometry of field-space can (and should) be studied using the cohomological properties of $\delta$ itself. 
Before closing this section, we make a couple of further remarks about these issues.

  First of all, the fact that $\delta\Sigma\neq0$, means that the explicit calculation of the presymplectic form $\Omega_H$ will likely become quite involved in any practical application. This fact might suggest the additional requirement\footnote{We used a hatted equality to mean that this is equation correspond to an extra conditions, and does not automatically follow from the framework.} 
\be
\delta_H\Sigma \,{\hat =}\, 0,
\label{sigma_var}
\ee 
so that $\Omega_H  \,{\hat =}\, \int_\Sigma\delta_H\theta_H$.
This condition states that the allowed \textit{physical} variations of the fields are only those which do not displace the hypersurface $\Sigma$.  
It can be interpreted as a condition selecting the physically viable boundary (and ``corner'') conditions to the problem under study. A more pictorial rephrasing---intentionally echoing the language of quantum mechanics---could be the following: given a finite and bounded region $\Sigma$ defined by the fields and to be interpreted as an observer's apparatus, then the only viable variations of the fields are those that leave this apparatus unchanged. It is the \emph{fixed frame} used to make our measurements. 

Finally, even supposing that the condition \eqref{sigma_var} is satisfied in a simple way, as for the Yang--Mills theory, our symplectic form generally does not reproduce---to the best of our knowledge---any other proposal advanced in the literature. This is again because of the non-vanishing of the functional curvature, $\frak F \neq 0$. Indeed, each time a $\delta_H^2$ appears in\footnote{We remind the reader that generically there is {\it no} local density $\zeta$ such that $\theta_H$ is equal to $\delta_H \zeta$.} 
\be
\Omega_H  \,{\hat =}\, \int_\Sigma\delta_H\theta_H,
\ee 
it gives a term (functionally) linear in the curvature $(2,0)$--form $\frak F$. The other terms will reproduce the usual presymplecitc form of general relativity, up to the replacement $\delta \leadsto \delta_H$:
\be
\Omega_H(g,\delta g, \varpi)  \,{\hat =}\, \Omega(g,\delta_Hg ) + ({\frak F}-\text{terms}).
\ee
Notice that the ${\frak F}-\text{terms}$ are purely corner terms on-shell of the equations of motions (cf \eqref{eq_komar}).
If we furthermore ignore the Gribov problem (see section \ref{sec:Gribov}), and require $\frak F$ to vanish, then again we can define a flat $\varpi$ connection {\it in an extended field space} as $\phi^{-1} \delta\phi$, where $\phi\in\mathrm{Diff}(\Sigma)$ (on-shell, $\phi\in\mathrm{Diff}(\partial\Sigma=C)$ suffices). In this situation, the ${\frak F}-\text{terms}$ disappear and we recover the presymplectic form proposed in \cite{Donnelly_2016} for general relativity, too. 

In the discussion session, we will expand on the physical interpretation of these remarks.

                                              
 \section{Geometric BRST \label{sec_brst}}
 
In this section, we comment on the relationship between the functional connection $\varpi$ and a geometric formulation of the BRST framework.\footnote{See e.g. \cite{Cotta-Ramusino, Cotta-Ramusino-Reina} and references therein, as well as \cite[chap. 8]{Bertlmann}.}
 In the usual formulation, the BRST (Slavnov) operator $s$ is a purely vertical variation (i.e. our $\delta_V$), and ghosts are vertical one forms, satisfying the Maurer--Cartan equation. The relationship to our construction can be summarized as follows: the BRST operator $s$ can be identified with the purely \textit{vertical} variation, $\delta_V=\fI_{\hat V}\delta$, while ghosts $\eta$ can be identified with $\varpi$ (not necessarily restricted to the vertical bundle). The standard interpretation is then recovered by assuming that a horizontal section in field-space is given, and that the (flat) functional connection $\varpi$ is fixed accordingly. Notice that, in the standard interpretation, the ghost one-form can be intrinsically defined only on the vertical vectors (   constituting the so-called Chevalley-Eilenberg complex \cite{Cotta-Ramusino-Reina}), while from our perspective this not only constitutes a superfluous restriction, but also an incomplete viewpoint. In fact, it can not fully account for the interplay between purely gauge and generic field variations. 
In order to justify our claims, we are going to briefly review the BRST framework for Yang--Mills theories.

In Landau gauge, the Yang--Mills Euclidean path integral can be written as
\be
\mathscr{Z}_\text{YM} = \int_{\cal F} \mathscr{D}\scrA \left| \det\left( \mathcal{M}(\scrA) \right) \right|\delta( B[\scrA] ) {\rm e}^{-S_\text{YM}},
\label{YM_path_integral} 
\ee
with $B[\scrA] = 0 $ the Landau gauge-fixing condition,
\be
B[\scrA] = \partial^\mu \scrA_\mu,
\label{Landaugauge}
\ee
and $\mathcal{M}(\scrA)$ the Faddeev--Popov ``matrix'',
\be
\mathcal{M}(\scrA)(x,y) = \frac{\delta }{\delta X(y)}_{|X=0}B[\scrA^X](x),
\ee
defined by the derivative of $B[\scrA]$ along an infinitesimal gauge transformation $X\in C^\infty(M)\otimes\frak g$, 
\be
\scrA^X := \scrA - \D_\scrA X.
\ee
It is not difficult to show that the FP determinant has an explicitly geometrical origin: 
it is the Jacobian appearing in the change of measure on $\cal F$ when a change of coordinates is performed adapted to the decomposition of field space along directions parallel to a section and to the orbits transversly intersecting it \cite{Jaskolski}.
The ghosts themselves are merely  a tool to calculate the determinant---as we  will now recall--- whereas BRST symmetry is essential, and  ensures independence with regards to the  choice of a section. 

Following Faddeev and Popov, the determinant in \eqref{YM_path_integral} 
can be re-written as an integration over anti-commuting variables $\eta$ and $\bar \eta$, named ghosts and anti-ghosts respectively: 
\be
\det\left(\mathcal{M}(\scrA)\right)=\int \mathscr{D}\eta  \mathscr{D}\bar\eta\;{\rm e}^{ -\int_M \bar\eta\cdot \mathcal{M}(\scrA)\cdot \eta} =: \int \mathscr{D}\eta  \mathscr{D}\bar\eta\;{\rm e}^{ - S_\text{FP}}.
\ee
where the dots correspond to DeWitt contraction (over internal indices and integration over spacetime points). 
We can then rewrite the path integral by replacing 
\be
S_\text{ YM}\rightarrow \tilde S_\text{YM} = S_{\text{ YM}}+S_{\text{ FP}}+S_{\text{ GF}},
\ee
and integrating over all the relevant variables. In this notation, $S_\text{FP}$ is the Faddeev-Popov term introduced above, while $S_{\text{GF}}$ corresponds to a rewriting of the gauge-fixing condition through a functional exponential, which usually involves some sort of Lagrange multiplier (which we won't discuss).  

Crucially, the total Lagrangian $\tilde L_\text{YM}$ turns out to be in the kernel of the {\it nilpotent} operator $s$, whose action on the Yang--Mills connection and ghosts variables is defined by
\be
\label{YM_BRST}
s \scrA= D_\scrA \eta \qquad \text{and}\qquad s \eta= -\frac12 [\eta, \eta] .
\ee

The second equation of \eqref{YM_BRST} is a Maurer--Cartan equation. As we have mentioned above, different authors have formally shown, from different perspectives, that $s$ has the interpretation of a vertical exterior derivative in field space, with $\eta$ playing the role of a functional Maurer-Cartan 1-form  of vertical variations. The issue, which is largely ignored, is that to relate a functional exterior derivative along the fibers with a general functional exterior derivative, one requires a splitting of the tangent space  $T{\cal F}$ (as a principal fiber bundle).  In turn, such a splitting is usually implied by the choice of a (functional) section of $\cal F$. 

 For instance, \textit{choosing a section} $A(x)$ as a representation of the equivalence class $[\scrA]$, we have the field-dependent gauge-transformed configuration:
\be 
A^\beta=\beta^{-1}(A(x)+\d)\beta(x) \in \mathcal{F}.
\ee
Varying, we obtain
\begin{align}
\delta A^\beta&=\delta \beta^{-1}(A+\d)\beta+\beta^{-1}\delta A\beta-\beta^{-1}A\delta \beta+\beta^{-1}\delta(\d\beta)\notag\\
&= -\d(\beta^{-1}\delta \beta)-A^\beta\beta^{-1}\delta \beta-\beta^{-1}\delta \beta \scrA+\beta^{-1}\delta A \beta\notag\\
&=-\D_{A^\beta}(\beta^{-1}\delta \beta)+\beta^{-1}\delta A \beta
\end{align}

 The second term represents a variation along the section. By setting it to zero, i.e. by considering a vertical restriction, we recover the Maurer--Cartan form: 
\be\label{MC} \delta_VA^\beta:=\fI_{\hat V}\delta A^\beta=-\D_{A^\beta}\eta
\ee
where here $\eta=\beta^{-1}\delta_V \beta$ is of the usual form. Note however, that not only is this solely defined for the vertical variation, but it also relies on the existence of the gauge section.

In the framework put forward in this note, the splitting issue can be easily overcome  without the requirement either of a section or of  zero curvature, since the splitting is precisely the responsibility of a general connection form. In fact, by comparing equation \eqref{YM_BRST} with equations \eqref{BRST_omega2} (up to a sign, see footnote \ref{fntminus}) and \eqref{deltaHA}, and recalling that $\delta_H := \delta - \delta_{\hat V}$, we see that the BRST operator $s$ can, indeed, be identified with $\delta_{\hat V}$, but $\eta$ need not be a Maurer--Cartan 1-form, since it can be identified with the full functional connection\footnote{Notice, however, that once a section is given, $\varpi$ can be fixed accordingly to be precisely of the form $\beta^{-1}\delta_V\beta$.}  $\varpi$. Hence, this setup liberates $\eta$ from being interpreted as a strictly flat connection.

With this interpretation in mind, the condition $s\, \tilde L_\text{YM}=0$   means that $\tilde L_\text{YM}$ is fully gauge invariant, in spite of the fact that it is no longer gauge invariant for a transformation involving solely the Yang--Mills fields (because of $S_\text{GF}$). Far from being something peculiar, this is a characteristic of any coupled system.

Finally, although it might look inconsistent that a {\it functional} 1-form, i.e. $\eta$, appears in the action, remember that ghosts $\eta$ always appear contracted with anti-ghosts $\bar \eta$.\footnote{Physical quantities should always be looked for among those of vanishing ghost-degree. In this counting ghosts and anti-ghosts have degree $\pm1$, respectively.} Therefore, a natural answer to this query is that anti-ghosts should be interpreted as functional vector-fields. In this manner,  the contraction of ghosts and anti-ghosts, i.e. of functional 1-forms and vector fields, would result in a standard scalar functional. We leave for future work further explorations in this direction,  leading to the full BV/BFV framework\footnote{The acronym stands for Batalin--Vilkovisky/Batalin--Fradkin--Vilkovisky.} \cite{BV, Cattaneo2014}.


\section{The Gribov problem\label{sec:Gribov}}

In finite-dimensional theories, the gauge connection can be used efficiently to probe topological properties of manifolds, as in e.g. Chern cohomology. Here, we advocate, but do not fully pursue, a similar approach to probing topological properties of {field} space through the use of functional connections. Indeed, there are at least two  appearances of   such properties in the standard lore of quantum field theory. 
   
Namely, both the Gribov problem and anomalies are allowed to arise from global topological properties of the principal fiber bundle. 
Whereas anomalies require a non-trivial first cohomology of the bundle (usually stated as ``ghost number 1'' cohomology of the BRST differential \cite{Bertlmann}),    the Gribov problem arises from a breakdown of the gauge-fixing  section \cite{gribov1977instability}. That is, from the non-triviality of the bundle itself. This breakdown could come from the fact that gauge-orbits intersect the orbit more than once, or, more severely, from the section becoming tangent to the orbits.
The latter more severe, since the inverse Faddeev--Popov determinant reaches a zero. Let us sketch how this comes about in Yang--Mills theory.

 In the Landau gauge \eqref{Landaugauge}, the Faddeev-Popov matrix requires the operator $\tilde{\mathcal M}$ to have a trivial kernel,

\be\label{invert_FP}
\tilde{\mathcal{M}}\xi  := \d\ast\D_\scrA \xi
\ee
with $\xi$ is a smooth Lie-algebra-valued function, and $\ast$ the spacetime Hodge operator.  Any element in the kernel of this operator represents an incomplete gauge-fixing, leaving directions in field space for which the usual propagator is degenerate. In the Euclidean case, for ``small enough'' $\scrA$, the operator is indeed invertible, as the Laplacian $(\d\ast\d)$ has a trivial kernel. However, in the non-perturbative regime, $\tilde{\mathcal M}$ will eventually develop zeros, since the term involving $\scrA$ in \eqref{invert_FP} is not of definite sign. This means that the section has to become, at some point, tangential to the orbits. 

This failure is not intrinsic to any particular choice of gauge-fixing, and can be obtained  from very general arguments \cite{singer1978some}. Let us sketch them here. Disregard the reducible gauge potentials $\scrA$, and the ones that have non-trivial stability group,\footnote{The original argument in \cite{singer1978some} does not require these assumptions, but its core is simplified with them.} and call $\mathcal{F}$ the space of the remaining gauge potentials. Then, $\mathcal{F}$ is a contractible space, i.e. within $\cal F$ one can deform any $n$-dimensional sphere (by which we mean a map $S^n\rightarrow \mathcal{F}$) to a point, and thus all of its homotopy groups are trivial, $\pi_j(\mathcal{F})=0$ for all $j\in\mathbb N$. The same is true for the image of a given section $\sigma:\mathcal{F}/\mathcal{G}\rightarrow \mathcal{F}$. Now, if ${\cal F}$ decomposed into a product $\text{Im}(\sigma)\times \mathcal{G}$, where $\mathcal{G}$ is the group of gauge transformations by the Lie group $G$,  then this would imply that  $\pi_j(\mathcal{G})=0$ for all $j$.  However, as it can be shown, for some $j\neq 0$, and underlying spacetime manifold given by $S^4$ (or $S^3$), and $G=\text{SU}(n)$, $n\geq 2$, there always exists a $j$ for which $\pi_j(\mathcal{G})\neq 0$. Therefore, the space $\mathcal{F}$ can {\it not} be generally decomposed into a product, and therefore there exists \emph{no global section} $\sigma:\mathcal{F}/\mathcal{G}\rightarrow \mathcal{F}$.

What is crucial to our discussion is that this result implies that there exists \emph{no} flat connection on $\mathcal{F}$. For, if there was one, the horizontal distribution would be integrable by the Frobenius theorem, and it would itself form a section.  
 
 The standard way of avoiding the Gribov problem is through the Gribov-Zwanziger  formalism \cite{Zwanziger}. Interestingly, its implementation requires the use of a non-standard BRST transformation. Since the standard BRST transformations correspond to  purely vertical variations, an interesting project would be to understand the geometrical meaning of these modified transformations. Our discussion seems to suggest that they could probe the curvature of $\varpi$.

 
\section{Discussion\label{sec:discussion}}
\paragraph*{Ghosts.}
The role of ghosts and BRST symmetry in the treatment of gauge field theories is well documented. Less widely known, but more elegant, is the geometric interpretation of such ghosts: if field space has the structure of a principal fiber bundle carrying an action of the group of gauge transformations, ghosts can be geometrically interpreted as  as Maurer--Cartan forms  along the gauge direction (see section \ref{sec_brst}). 

In particle theory however, ghosts move only 'behind the scenes', never showing up on external legs of Feynman diagrams---legs which represent asymptotic particle states. When glued together to form inner loops, previously external legs no longer correspond to asymptotic states, and again ghosts are required to ensure gauge-invariance {and unitarity} of the amplitude.\footnote{This is literally true in non-Abelian gauge theories. For Abelian theories, gauge-transformations are field-independent (don\rq{}t depend on the base point of $P$),  making the use of connections less important and integration over gauge degrees of freedom easily factorable.}

\paragraph*{Boundaries and gauge-symmetries.}
However, it is  difficult to compare the role of gauge transformations in finite bounded regions and  asymptotically---where gauge freedom is frozen. Indeed, recently the importance of non-asymptotic boundaries and corners in field theories has undergone further scrutiny, {\it viz \'a viz}  gauge-invariance \cite{Donnelly_2016}  (but see also \cite{Cattaneo2014}). These recent investigations found that to keep gauge-covariance for subsystems---subsystems which can in principle be reconnected like legs of the Feynman diagram---one requires the addition of further terms in the symplectic potential of the theory. 

Also, note that the total charge associated to a local gauge symmetry must vanish (on-shell). Repackaged, this means solely that gauge is a ``redundancy'' encoded in a constraint, rather than a symmetry related to a conserved charge. In the presence of boundaries, however, the situation becomes more obscure: gauge transformations with support on the boundary give rise to non-vanishing charges, and  whether these gauge transformations should be dealt with as pure redundancies or as actual symmetries becomes questionable. Exploration of this question has begun in an influential series of papers,  \cite{He:2014cra,He:2015zea,He:2014laa,Barnich2010,Campiglia2014,Campiglia2016}.

\paragraph*{Boundaries, ghosts, and the connection-form.}
Here we have connected these two dots by the use of the connection-form. This was done by showing that the necessary terms for implementing gauge-covariance at finite boundaries come precisely from a generalization of the interpretation of ghosts as forms in field space. To achieve this, we constructed a geometrical theory of field-space-dependent gauge-covariance, complete with a general field-space connection $\varpi$ and associated curvature $\frak F$. 

In particular, we advocated the identification of ghosts $\eta$ with the functional connection $\varpi$. But, were $\varpi$ identified with a Maurer--Cartan form, its curvature would be forced to vanish. Importantly, however,  our framework does {\it not} require the functional curvature $\frak F$ to vanish. In fact, the Maurer--Cartan equation has to hold for vertical variations only, and---as already observed, cf. equation \eqref{BRST_omega2}---this holds irrespectively of the vanishing of $\frak F$, and rather follows from its horizontality. Therefore, while encompassing the usual geometrical BRST framework, our setting generalizes it.
Notably,  this generalization is actually necessary for a consistent treatment of gauge theories in field space. That is because, due to the Gribov problem, field space cannot have a globally vanishing functional curvature \cite{singer1978some}, nor can it have a global section from which to separate physical variables from gauge directions. To stress this point,  \emph{the more general formalism developed here---which does not rely on sections and allows for non-vanishing curvature---is necessary for a global treatment of gauge-invariance in field space.}

\paragraph*{Path integral, functional renormalization, and boundaries.}
Even formally writing a non-perturbative path integral in the presence of Gribov ambiguities is no easy task. And ensuring gauge-invariance non-pertur-batively is similarly non-trivial. In the background-field formalism, one needs to ensure \lq\lq{}split-symmetry\rq\rq{}---the one corresponding to the arbitrariness in how one splits the field into background and perturbations, $g=g_o+h$---and hence use Nielsen identities to ensure that diffeomorphism symmetry is respected by the computation and the split throughout. 

In order to avoid some of these difficulties in the non-perturbative functional renormalization group (FRG) approach, a fully diffeomeorphism--invariant implementation of the FRG was put forward which used the Vilkovisky connection and the Vilkovisky-DeWitt effective action formalism by Pawlowski {\it et al.} \cite{Pawlowski1,Pawlowski2}.
Their construction can be understood as, in their words, \lq\lq{}a non-linear upgrade of the standard background field approach, its linear order giving precisely the background field relations in the Landau-DeWitt gauge. The gain of such a non-linear approach is that the fluctuation fields have a geometrical meaning and can be utilised to compute an effective action which only depends on the diffeomorphism-invariant part of the fluctuation fields. Consequently, the geometrical effective action is trivially diffeomorphism-invariant, and any cutoff procedure applied to these fluctuation fields maintains diffeomorphism invariance\rq\rq{}.
Thus, although it is not obvious that  employing  such functional geometrical methods will give results which diverge from more standard ones, it is clear that they are indeed the most efficient ones to explicitly ensure gauge-covariance in the non-linear regime. 

Moreover, it seems to us that in the presence of boundaries, since the functional connection becomes indispensable already at the classical  level, it is to be expected that indeed differing results will ensue, depending on whether one computes quantities using the connection-form or not (and on the choice of connection form as well).%
\footnote{In the authors opinion,  it is also questionable whether one can fully make sense of gauge in presence of boundaries without using the functional connection.}

Pawlowski {\it et al}'s work illustrates the potential such connection forms have in the FRG approach. However, their work utilizes exclusively the Vilkovisky connection, as does most other previous work implicitly involving connection-forms. It is thus conceivable that the generalization our work allows in terms of new choices of connection may have computational gains. 

\paragraph*{A physical interpretation}
In physical terms, the role of the functional connection $\varpi$ is---as we have repeatedly emphasized---that of establishing which part of generic field variation is gauge and which part is ``physical''. Telling apart these two contributions requires a splitting of $\mathrm T\mathcal F$ into vertical and horizontal directions. Such a splitting is provided, e.g., by a full-fledged field-space connection $\varpi$, satisfying the properties \eqref{connection1and2}. Notice that gauge transformations will simultaneously transform the fields as well as the connection $\varpi$.  In the usual particle theory language, it is the ensuing cancellation from these transformations that would get interpreted through the Kugo--Ojima mechanism.

{Furthermore, we also showed that the use of horizontal---i.e. gauge-covariant---variations is a natural and physically sound way to compute a theory's gauge-invariant presymplectic potential from its action functional. In this manner, the usual presymplectic potential gets shifted by terms containing the functional connection $\varpi$ and its curvature. Under extra assumptions, these terms reduce to those advocated by \cite{Donnelly_2016}.
Those authors interpreted these extra terms as new ``boundary'' degrees of freedom, while our construction sees them in a different light:
we do not need to \emph{enlarge} the space of fields in an {\it ad hoc} way; we rather endow it with extra geometric structures. In fact, one could argue they should have been present from the onset---our $\varpi$ is not a field {\it on spacetime} but encodes the geometry {\it of field-space}.

The presymplectic potential depends on the choice of a hypersurface, $\Sigma$, and its calculation requires extra care when dealing with background independent theories. In Yang--Mills theory, $\Sigma$ can be assumed fixed {\it a priori} on the spacetime manifold, without interfering with gauge symmetries. In general relativity, however, the transformation properties of the presymplectic potential density demand that the surface should itself be defined relationally, i.e. in terms of the dynamical fields, which means $\delta\Sigma\neq0$. Nevertheless, our covariant formalism allows us to implement the auxiliary condition $\delta_H\Sigma=0$, largely simplifying computations.  This condition has a neat physical interpretation: it posits $\Sigma$ to be determined by fields whose physical---i.e. horizontal---variations vanish at $\Sigma$. It is a restriction to regions whose boundaries suffer no physical variation themselves. This seems like a reasonable requirement of a ``good frame of reference'' for measurements.

Up to this point, we have mostly discussed the geometric structure of field space. However, apart from geometric consistency, the connection form still seems very much undetermined. Any choice made with the correct transformation laws appears to be equally good. What selects specific connection forms, then?

We would like to put forward the following tentative interpretation: that it be related to a choice of observer.\footnote{%
Such a proposal is in many aspects analogous to that of \cite{Donnelly_2016}, although it is here made more encompassing and precise. There is also an overlap of ideas with \cite{whygauge}.}
Interpreting $\varpi$ as an observer fits with the fact that a choice of $\varpi$ is needed to tell apart physical and ``fictitious'', i.e. pure-gauge, field changes.
In other words, it allows reference to ``objective change'' in gauge theories and general relativity,  objective at least with respect to an (abstract) observer. In fact, suppose we are given two different, infinitesimally close field configurations $g_{ab}$ and $g_{ab}+ \gamma_{ab}$, on two subsequent (infinitesimally close) hypersurfaces $\Sigma_g$ and $\Sigma'_{g+\gamma}$; then, once a functional connection $\varpi$ is fixed, i.e. once an observer has been chosen, we can say how much of $\gamma_{ab}$ is pure gauge, that is $\hat V (\gamma) = (\varpi(\gamma))^\#$, and how much is an actual variation with respect to the observer, $\hat H(\gamma) = \gamma - \hat V(\gamma)$.

Hence, the functional connection produces an ``identification vector-field'' $\varpi(\gamma)$ between the points on the two hypersurfaces, depending on how the field is infinitesimally changing between them.\footnote{In this restricted sense, it parallels the constructions of \lq\lq{}best-matching\rq\rq{} vector fields in gravity \cite{Barbour_RWR} (or \cite{Flavio_tutorial} and references therein). See \cite{Gauge_Riem} for an explanation of best-matching in terms of a fixed connection one-form in the canonical gravitational setting. See also \cite{Littlejohn} for an application of such connection forms for the rotation group in molecular dynamics.} In the gauge-theory setting, this identification would be among internal spaces corresponding to the two sides of the hypersurface.
Of course, points on manifolds (or equivalently, points along an internal fiber) do not make sense by themselves, and their identification must be made with respect to the field content and embody equivariance. Even in the presence of material reference frames, however, many choices can be made, within the constraints set by  equivariance laws. In other words, any choice can be made, as long as it be given by a connection form, i.e. by what we want to call \textit{an abstract observer}.

Work in progress already shows that this abstract observer could be chosen so as to follow a physical field. More abstractly, it could be given by \lq\lq{}Einstein's mollusk\rq\rq{}---representing an actual material reference frame (one whose equations of motion are possibly unknown).
But the connection $\varpi$ is not itself physical, as it does {\it not} necessarily belong to the space of fields. 
The introduction of an {\it abstract} observer can be particularly fecund, because it liberates us from the burden of dealing with the observer's dynamics and back-reaction. This is analogous to the quantum mechanics' observer, which is rarely modeled as an actual physical device, and is most often understood as the choice of a complete set of mutually commuting Hermitian operators. 

The need of identifying points on manifolds or internal spaces emerges in particular when considering stitching patches of space(time) to one-another.
And it is precisely this stitching which makes the requirement of the functional connection (ghosts) manifest for physics. Without this requirement, we can loose unitarity, since certain vertical field transformations will not have the accompanying transformations of the stitching, and will be ``counted as physical''. 

In the light of the interpretation of $\varpi$ as an abstract observer, it is also interesting to reexamine  the meaning of the condition $\delta_H\Sigma=0$, which emerged in the study of (pre)symplectic geometry within the hypersurface $\Sigma$. According to this interpretation, the condition $\delta_H\Sigma=0$ says that the boundary surface $\Sigma$ has to be determined by those fields whose {\it physical variations with respect to the observer} (``horizontal variations with respect to $\varpi$'') vanish at $\Sigma$. This is particularly compelling not only because it implicitly contains many of the ideas we have so far presented, but also because it offers a further bridge with the quantum mechanics notion of observer. Let us explain how this comes about. First of all, $\delta_H\Sigma=0$ exemplifies the relationship between observers and boundary interfaces: finite and bounded regions are in fact the natural loci of measurement apparata. On top of that, it draws a line between the study of gauge-invariant (pre)symplectic geometry on a hypersurface $\Sigma$ and the relational specification of $\Sigma$ itself.

\section*{Acknowledgment}

 We would like to thank Ant\^onio Duarte for having read the manuscript and provided us with useful comments.
This research was supported in part by Perimeter Institute for Theoretical Physics. Research at Perimeter Institute is supported by the Government of Canada through the Department of Innovation, Science and Economic Development Canada and by the Province of Ontario through the Ministry of Research, Innovation and Science.

 \appendix
 \section{Diffeomorphism, vector fields, and Lie derivatives \label{app}}

To make matters more clear, in this appendix we consider two a priori distinct manifolds $M, N$, and a diffeomorphism
\be
\varphi:M\rightarrow N.
\ee
Similarly, we consider $\mathcal{F}$ and $\tilde{\mathcal{F}}$ the spaces of metric fields over $M$ and $N$, respectively.  Let $x\in M$ and $y\in N$, $g\in \mathcal{F}, \tilde g \in \tilde{\mathcal{F}}$, $f\in C(M)$ and $h\in C(N)$,  $\mu\in C({\cal F})$ and $\eta\in C(\tilde{\mathcal{F}})$, and finally $X\in\Gamma({\rm T}M)$ and $Y\in\Gamma({\rm T} N)$. 

If $\varphi:M\rightarrow N$ sends $x\in M$ to $\varphi(x)\in  N$, the pullback $\varphi^*:\tilde {\cal F} \to {\cal F}$ goes in the opposite direction. 
This defines the right--composing map
\be
A_{\varphi^*}:\tilde {\cal F} \to {\cal F}\;, \quad \tilde g \mapsto A_{\varphi^*}\tilde g = \varphi^* \tilde g.
\ee 
Its pullback maps the space of functionals in the opposite direction, i.e. the same as $\varphi$,
\be
A_{\varphi^*}^*:C({\cal F}) \to C(\tilde{\cal F}),\quad \mu \mapsto A_{\varphi^*}^*\mu = \mu\circ \varphi^*.
\ee
The infinitesimal version of a diffeomorphism is a vector field.
Define $\Psi_{(tY)}:N\to N$, with $\Psi_0 = \text{id}$, to be the flow of $Y$:
\be
\frac{\d}{\d t} \Psi_{(tY)} = Y \circ \Psi_{(tY)}.
\ee
The pushforward of a  vector field is given by the map
\be
\varphi_* : \Gamma({\rm T}M) \to \Gamma({\rm T}N)
\ee
so that
\be
[(\varphi_* X)h ]_{\varphi(x)} = [X (h\circ\varphi)]_{x} \,.
\label{papparappa}
\ee
Now, a vector field on $\tilde{\cal F}$ can be pushed-forward along $A_{\varphi^*}$ in a similar way.
\be
(A_{\varphi^*})_*: \Gamma({\rm T}\tilde {\cal F}) \to \Gamma({\rm T} {\cal F})
\ee
so that
\be
\Big[\Big((A_{\varphi^*})_* Y^\#\Big)\mu\Big]_{A_{\varphi^*} \tilde g} 
= \Big[ Y^\#\Big(\mu\circ A_{\varphi^*} \Big)\Big]_{\tilde g}
= \Big[ Y^\#\Big(\mu\circ\varphi^* \Big)\Big]_{\tilde g}
\ee
Now, recall equation \eqref{vertical_action},
\be
(Y^\# \eta)_{\tilde g} 
= \frac{\d}{\d t}_{|t=0} (A_{\Psi^*_{(tY)}}^*\eta)_{\tilde g}=\frac{\d}{\d t}_{|t=0}\eta_{\Psi_{(tY)}^*\tilde g} \; .
\ee
Thus
\begin{align}
\Big[\Big((A_{\varphi^*})_* Y^\#\Big)\mu\Big]_{A_{\varphi^*}\tilde g} 
& = \frac{\d}{\d t}_{|t=0} (\mu \circ \varphi^*)_{\Psi^*_{(tY)} \tilde g }
= \frac{\d}{\d t}_{|t=0} \mu_{\varphi^*(\Psi_{(tY)} )^* \tilde g }
= \frac{\d}{\d t}_{|t=0} \mu_{(\Psi_{(tY)}\circ\varphi )^* \tilde g }\notag\\
~&= \frac{\d}{\d t}_{|t=0} \mu_{(\varphi^{-1}\circ\Psi_{(tY)} \circ \varphi)^* \varphi^*\tilde g }
=\frac{\d}{\d t}_{|t=0}  \Big(A^*_{(\varphi^{-1}\circ\Psi_{(tY)} \circ \varphi)^*}\mu\Big)_{\varphi^* \tilde g}
\end{align}
Finally, $\frac{\d}{\d t} ( \varphi^{-1}\circ\Psi_{(tY)} \circ \varphi ) = X \circ \varphi^{-1}\circ\Psi_{(tY)} \circ \varphi $ defines a vector field $X$, which can be computed by comparing
\be
\frac{\d}{\d t}_{|t=0} ( \varphi^{-1}\circ\Psi_{(tY)} \circ \varphi )^*f_x =  \frac{\d}{\d t}_{|t=0} ( \Psi_{(tY)})^*(f\circ\varphi^{-1})_{\varphi(x)},
\ee
and the analogue of equation \eqref{papparappa}:
\be
\Big[\Big(\varphi^{-1})_* Y\Big) f\Big]_{\varphi^{-1}(y)} = \Big[Y(f\circ\varphi^{-1})\Big]_{y} = \Big[\pounds_Y(f\circ\varphi^{-1})\Big]_{y} 
= \frac{\d}{\d t}_{|t=0} \Psi_{(tY)}^*(f\circ\varphi^{-1})_{y},
\ee
obtaining
\be
X = (\varphi^{-1})_* Y\circ \varphi.
\ee
Thus,
\be
((A_{\varphi^*})_* Y^\#)_{\varphi^*\tilde g} =((\varphi^{-1})_*Y\circ\varphi)^\#_{\varphi^*\tilde g} \;.
\label{crucial}
\ee
Clearly, this result could have been anticipated as the only reasonable one, at least once the source and target manifolds of $\varphi$, $M$ and $N$, have been distinguished and the direction of the maps made clear.

\section{Derivation of some fundamental equations}

\subsection{ Variation of an equivariant horizontal $(0,p)$-form\label{app2}}

We are going to show explicitly equation \eqref{eq_app2}, i.e. that for an equivariant $(0,p)$-form $\mu$
\be
\delta (\mu(\beta^*g))= \Big(\delta \mu +  \fL_{(\delta\beta\circ\beta^{-1})^\#}\mu \Big)(\beta^*g).
\ee
Indeed, 
\begin{align}
\delta ( A_{\beta_g^*}^*\mu)_{|g}
& =  A_{\beta_{g+\delta g}^*}^*\mu_{|g+\delta g} -  A_{\beta_g^*}^*\mu_{|g} \notag\\
& = A_{\beta^*_{g+\delta g}}^*\mu_{|g} + A_{\beta^*_{g}}^*\mu_{|g+\delta g} -  2 A_{\beta^*_g}^*\mu_{|g} \notag\\
& = A_{\beta^*_g}^* A_{(\text{id} + \delta\beta \circ \beta^{-1})^*_g}^*\mu_{|g} +  A_{\beta^*_{g}}^*(\mu + \delta \mu)_{|g}-  2 A_{\beta^*_g}^*\mu_{|g} \notag\\
& = A_{\beta^*_g}^*\Big( \fL_{(\delta\beta\circ\beta^{-1})^\#} \mu + \delta\mu  \Big) _{|_g}\notag\\
&=\Big( \fL_{(\delta\beta\circ\beta^{-1})^\#}\mu + \delta \mu \Big)_{|\beta^*_g g}
\end{align}
where, to pass from the second to the third line, we used that pullbakcs compose to the right, while the last step follows from the equivariance of $\mu$.

\vspace{1em}~\\

\subsection{Direct proof of equation \eqref{ymdeltaomega}\label{app_direct}}

We want to prove by means of direct calculation that $\Omega_H = \delta \Theta_H$.
We do this explicitly, to give an example of a calculation performed with $(n,p)$-forms.

We first prove a couple of useful identities. All equations should be understood within a trace, and products forms should be understood as wedge products in the appropriate space.
\begin{subequations}
\begin{align}
\varpi\delta \D_\scrA E & \stackrel{[\d,\delta]=0}{=} \varpi( \d \delta E + [\delta \scrA , E ] + [\scrA, \delta E]) \\
[\varpi, E] \d\varpi & \stackrel{\text{ trace}\atop\text{ad-invariance}\phantom{p}}{=} \varpi [E,\d\varpi]  \stackrel{{\text{trace cyclicity \&} \atop \text{commuting}}\atop \text{$(1,0)$-forms}}{=} - [E,\d\varpi] \varpi \stackrel{\text{ trace}\atop\text{ad-invariance}\phantom{p}}{=} -E[\d\varpi,\varpi]  \stackrel{[\d,\delta]=0}{=}  -\frac12 E\d[\varpi,\varpi] \\
[\varpi, E][\varpi, \scrA] & \stackrel{\text{ trace}\atop\text{ad-invariance}\phantom{p}}{=} [ [\varpi, E],\varpi]\scrA \stackrel{{\text{commutator} \atop \text{antisym. \&}}\atop{\text{commuting}\atop \text{$(1,0)$-forms}}}{=} [ [\varpi ,E],\varpi]\scrA  \stackrel{\text{graded}\atop\text{Jacobi id.}}{=}\frac12[[\varpi,\varpi],E]\scrA = \stackrel{\text{ trace}\atop\text{ad-invariance}\phantom{p}}{=} \frac12 [\varpi,\varpi] [E,\scrA]
\end{align}
\end{subequations}
With these equations at hand, via similar manipulations we find
\begin{align}
 \delta_H E\delta_H \scrA 
& = \delta E \delta A - [\delta E, \scrA]\varpi  + \delta E \d\varpi  + \varpi[E ,\delta \scrA] + [\varpi, E][\varpi, \scrA] - [\varpi, E]\d\varpi\notag\\
& =  \delta E \delta A + \Big(  [\scrA, \delta E]\varpi  + (\d\delta E) \varpi  + [\delta \scrA, E]\varpi \Big) -\d((\delta E) \varpi)+ \frac12 [\varpi,\varpi] [E,\scrA]+ \frac12 E\d[\varpi,\varpi]\notag\\
& = \delta E \delta A + (\delta\D_\scrA E)\varpi  -\d\delta (E \varpi)  + \d(E\delta\varpi)+ \frac12 [\varpi,\varpi] [E,\scrA] - \frac12 (\d E)[\varpi,\varpi] + \frac12 \d( E[\varpi,\varpi])\notag\\
& =\delta E \delta A + \delta (\varpi \D_\scrA E ) -(\D_\scrA E)\delta\varpi -\d\delta (E \varpi) -  \frac12 (\D_\scrA E)[\varpi,\varpi] + \d( E{\frak F})\notag\\
& =\delta E \delta A + \delta (\varpi \D_\scrA E )  -\d\delta (E \varpi) -(\D_\scrA E){\frak F} + \d( E{\frak F})\notag\\
& =\delta E \delta A + \delta (\varpi \D_\scrA E )  -\delta \d(E \varpi) + E\D_\scrA{\frak F} 
\end{align}
And finally, 
\begin{align}
\Omega_H = \frac12 \int_\Sigma \delta_H E\delta_H \scrA - E\D_\scrA {\frak F} =  \frac12 \int_\Sigma \delta E \delta\scrA + \delta(\varpi \D_\scrA E) - \delta (\d E\varpi) = \delta \Theta_H.
\end{align}

  
\bibliographystyle{bibstyle_aldo}


\end{document}